# Alternating Optimization for Capacity Region of Gaussian MIMO Broadcast Channels with Per-antenna Power Constraint

Thuy M. Pham, Ronan Farrell, and Le-Nam Tran
Department of Electronic Engineering, Maynooth University, Maynooth, Co. Kildare, Ireland
Email: {minhthuy.pham, ronan.farrell, lenam.tran}@nuim.ie

*Abstract*—This paper characterizes the capacity region of Gaussian MIMO broadcast channels (BCs) with per-antenna power constraint (PAPC). While the capacity region of MIMO BCs with a sum power constraint (SPC) was extensively studied, that under PAPC has received less attention. A reason is that efficient solutions for this problem are hard to find. The goal of this paper is to devise an efficient algorithm for determining the capacity region of Gaussian MIMO BCs subject to PAPC, which is scalable to the problem size. To this end, we first transform the weighted sum capacity maximization problem, which is inherently nonconvex with the input covariance matrices, into a convex formulation in the dual multiple access channel by minimax duality. Then we derive a computationally efficient algorithm combining the concept of alternating optimization and successive convex approximation. The proposed algorithm achieves much lower complexity compared to an existing interior-point method. Moreover, numerical results demonstrate that the proposed algorithm converges very fast under various scenarios.

## I. Introduction

Since its invention in the mid-90s [1], [2], multiple-input multiple-output (MIMO) technology has been adopted in all modern mobile wireless networks. From a system design perspective, one of the most fundamental problems is to compute the capacity of the system of interest. For a single user MIMO (SU-MIMO) channel, pioneer studies proved that the capacity can be achieved by Gaussian input signaling [1], [2]. For multiuser MIMO scenarios, the seminal work of [3] showed that dirty-paper coding (DPC) in fact achieves the entire capacity region of Gaussian MIMO broadcast channels (BCs). The particular case of the sum capacity of MIMO BCs was studied in several pioneer studies including [3]–[7],

The capacity of MIMO systems is investigated along with a certain type of constraint on the input covariance matrices. In this regard, we remark that all papers mentioned above assume a sum power constraint (SPC), and this usually leads to efficiently computational algorithms. For the sum-capacity computation, Viswanathan *et al.* [8] applied a steepest descend method, while Yu [9] proposed a dual decomposition-based algorithm. In this line of research, Jindal *et al.* presented a sum power iterative water-filling algorithm by exploiting the MAC-BC duality. The entire capacity region of MIMO BCs with a SPC was characterized in [10], [11], using conjugate gradient projection (CGP)- and pre-conditioned gradient projection-based approaches.

While SPC has been widely considered, it is less appealing in reality due to the fact that each antenna is usually equipped with a different power amplifier, which has its own power budget. Despite its practical and fundamental importance, the research on efficient methods for computing the capacity region of Gaussian MIMO BCs has been quite limited. For SU-MIMO, this problem was solved in [12], [13] by the so-called mode-dropping algorithm. In [3], it was shown that DPC still achieves the full capacity region of the MIMO BC under PAPC. However, finding the DPC region with PAPC is more numerically difficult than with a SPC. In fact, no closed-form design has been reported for the computation of the capacity region of the MIMO BC subject to PAPC. To the best of our knowledge, the only attempt to characterize the entire capacity region of the MIMO BC subject to PAPC was made in [14]. Specifically, the authors in [14] established a modified duality between the MAC and BC and transformed the input optimization problem in the BC into a minimax optimization problem in the dual MAC. Then resulting program is solved by a standard barrier interior-point routine. Thus, as a common property for the class of second order optimization methods, the algorithm proposed in [14] has computational complexity that does not favor large-scale antenna systems which are envisioned in next wireless communications generations.

In this paper, we determine the capacity region of MIMO BCs under PAPC. In particular, the problem of interest is also known as weighted sum rate maximization (WSRMax) for MIMO BCs. As mentioned earlier, the capacity region can be achieved by DPC, but the resulting WSRMax problem is nonconvex. As a standard step, we apply the minimax duality presented in [14] to transform the WSRMax problem into a minimax program. However, unlike [14] which solves the resulting minimax problem by a barrier interior-point method, we take advantage of the problem specifics to propose an efficient algorithm that blends the concept of alternating optimization (AO) and successive convex approximation (SCA). Especially, in each iteration of the proposed algorithm, closed-form expressions based on conjugate gradient projection (CGP) method are derived. As a result, the complexity of the proposed algorithm is much lower than that of the barrier method in [14], and scales linearly with the number of users in the system, making it particularly suitable for large-scale networks. Numerical experiments are carried out to demonstrate that the proposed algorithm can converge rapidly, especially for networks of high signal-to-noise ratio, and the number of iterations required for convergence is quite insensitive to the number of users.

The remainder of the paper is organized as follows. The system model is described in Section II followed by the proposed algorithm in Section III. Section IV provides the complexity analysis of the proposed algorithm while Section V presents the numerical results. Finally, we conclude the paper in Section VI.

*Notation*: Standard notations are used in this paper. Bold lower and upper case letters represent vectors and matrices, respectively. $\mathbf{I}$ defines an identity matrix, of which the size can be easily inferred from the context; $\mathbb{C}^{M \times N}$ denotes the space of $M \times N$ complex matrices; $\mathbf{H}^\dagger$ and $\mathbf{H}^T$ are Hermitian and normal transpose of $\mathbf{H}$, respectively; $\mathbf{H}_{i,j}$ is the $(i,j)$th entry of $\mathbf{H}$; $|\mathbf{H}|$ is the determinant of $\mathbf{H}$; $\text{rank}(\mathbf{H})$ stands for the rank of $\mathbf{H}$; $\text{diag}(\mathbf{x})$ denotes the diagonal matrix with diagonal elements being $\mathbf{x}$.

## II. SYSTEM MODEL

Consider a $K$-user MIMO BC where the base station and each user have $N$ and $M_k$ antennas, respectively. The channel matrix for user $k$ is denoted by $\mathbf{H}_k$. Let $\mathbf{s}$ be the composite signal that combines the data for all users in the downlink. Then, the received signal at user $k$ is expressed as

$$\mathbf{y}_k = \mathbf{H}_k \mathbf{s} + \mathbf{z}_k \quad (1)$$

where $\mathbf{z}_k$ is the Gaussian noise with distribution $\mathcal{CN}(\mathbf{0}, \mathbf{I}_M)$. When DPC is applied to achieve the capacity region, for a given user $k$, the interference caused by users $j < k$ is completely canceled without affecting the optimality. As a result, the WSRMax under PAPC is formulated as

$$\begin{array}{ll} \underset{\{\mathbf{S}_k \succeq \mathbf{0}\}}{\text{maximize}} & \sum_{k=1}^{K} w_k \log \left| \frac{|\mathbf{I}+\mathbf{H}_k \sum_{i=1}^{k} \mathbf{S}_i \mathbf{H}_k^\dagger|}{|\mathbf{I}+\mathbf{H}_k \sum_{i=1}^{k-1} \mathbf{S}_i \mathbf{H}_k^\dagger|} \right| \\ \text{subject to} & \sum_{k=1}^{K} [\mathbf{S}_k]_{i,i} \leq P_i, \forall i \end{array} \quad (2)$$

where $\mathbf{S}_k$ is the input covariance matrix for the $k$th user, $P_i$ is the power constraint on antenna $i$, and $w_k$ is the weighting factor assigned to user $k$. Without loss of optimality, we assume that $0 < w_1 \leq w_2 \leq ... \leq w_K$ and $\sum_{k=1}^{K} w_k = 1$ in the following. Since (2) is a nonconvex problem, solving it directly is not a good option. However, we can exploit the MAC-BC duality to transform (2) into minimax program in the dual MAC which can be solved efficiently by the novel AO as presented in the next section.

## III. PROPOSED SOLUTION

Applying the modified MAC-BC duality introduced in [15], we can equivalently rewrite (2) as the following minimax optimization problem

$$\begin{array}{ll} \underset{\mathbf{Q} \succeq \mathbf{0}}{\min} \underset{\{\bar{\mathbf{S}}_k \succeq \mathbf{0}\}}{\max} & \sum_{k=1}^{K} \Delta_k \log |\mathbf{Q} + \sum_{i=k}^{K} \mathbf{H}_i^\dagger \bar{\mathbf{S}}_i \mathbf{H}_i| \\ & -w_K \log |\mathbf{Q}| \\ \text{subject to} & \sum_{k=1}^{K} \text{tr}(\bar{\mathbf{S}}_k) = P, \\ & \text{tr}(\mathbf{QP}) = P, \mathbf{Q} : \text{diagonal} \end{array} \quad (3)$$

where $\Delta_k = w_k - w_{k-1} \geq 0$, $P \triangleq \sum_{i=1}^{N} P_i$; $\{\bar{\mathbf{S}}_k\}$ and $\mathbf{Q}$ are input covariance and noise covariance matrices in the dual MAC, respectively. As shown in [15], the objective in (3) is convex with $\mathbf{Q} \succeq \mathbf{0}$ and concave with $\{\bar{\mathbf{S}}_k \succeq \mathbf{0}\}$. Thus, there exists a saddle point for (3), which also solves (2). The minimax formulation in (3) also suggests a way to find $\{\bar{\mathbf{S}}_k\}$ and $\mathbf{Q}$ by AO. However, a pure AO algorithm is not guaranteed to converge. In fact, a counterexample was already given in [16]. In what follows, we propose an iterative algorithm based on combining AO and SCA, of which convergence can be proved.

Let $\{\bar{\mathbf{S}}_k^n\}$ be the solution to the following maximization problem in the $n$th iteration

$$\begin{array}{ll} \underset{\{\bar{\mathbf{S}}_k \succeq \mathbf{0}\}}{\text{maximize}} & \sum_{k=1}^{K} \Delta_k \log |\mathbf{Q}^n + \sum_{i=k}^{K} \mathbf{H}_i^\dagger \bar{\mathbf{S}}_i \mathbf{H}_i| \\ \text{subject to} & \sum_{k=1}^{K} \text{tr}(\bar{\mathbf{S}}_k) = P. \end{array} \quad (4)$$

The above maneuver is nothing but a standard routine of optimizing $\{\bar{\mathbf{S}}_k\}$ when $\mathbf{Q}$ is held fixed. Problem (4) can be solved by off-the-shelf interior-point convex solvers but the complexity is not affordable for large-scale systems. In our numerical experiments, all of known (free and commercial) solvers fail to solve (4) on a relatively powerful desktop PC for $N \geq 100$, regardless of the number of users. That is, interior-point methods are not an efficient approach to solving (4) for massive MIMO techniques which are likely to be adopted in 5G systems. To overcome this shortcoming, we now present an efficient method to solve (4) based on the CGP framework. To proceed, let $\mathcal{S} = \{\bar{\mathbf{S}}_k | \bar{\mathbf{S}}_k \succ \mathbf{0}, \sum_{k=1}^{K} \text{tr}(\bar{\mathbf{S}}_k) = P\}$ be the feasible set of (4). The main operation of a CGP method is to project a given $\{\tilde{\mathbf{S}}_k\}$ onto $\mathcal{S}$. Our motivation is that the projection of $\{\tilde{\mathbf{S}}_k\}$ onto $\mathcal{S}$ can be reduced to a projection of a resulting vector onto a canonical simplex, which can be computed efficiently.

The projection of $\{\tilde{\mathbf{S}}_k\}$ onto the feasible set $\mathcal{S}$ is formulated as

$$\begin{array}{ll} \underset{\{\dot{\mathbf{S}}_k \succeq \mathbf{0}\}}{\text{minimize}} & \sum_{k=1}^{K} ||\dot{\mathbf{S}}_k - \tilde{\mathbf{S}}_k||_F^2 \\ \text{subject to} & \sum_{k=1}^{K} \text{tr}(\dot{\mathbf{S}}_k) = P. \end{array} \quad (5)$$

Let $\mathbf{U}_k \tilde{\mathbf{D}}_k \mathbf{U}_k^\dagger = \tilde{\mathbf{S}}_k$ be the EVD of $\tilde{\mathbf{S}}_k$, where $\mathbf{U}_k$ is unitary and $\tilde{\mathbf{D}}_k$ is diagonal. Then we can write $\dot{\mathbf{S}}_k = \mathbf{U}_k \dot{\mathbf{D}}_k \mathbf{U}_k^\dagger$ for some $\dot{\mathbf{D}}_k \succeq \mathbf{0}$. Since $\mathbf{U}_k$ is unitary, it holds that $\text{tr}(\dot{\mathbf{S}}_k) = \text{tr}(\dot{\mathbf{D}}_k)$ and that $||\dot{\mathbf{S}}_k - \tilde{\mathbf{S}}_k||_F = ||\dot{\mathbf{D}}_k - \tilde{\mathbf{D}}_k||_F$. That is to say, (5) is equivalent to

$$\begin{array}{ll} \underset{\{\dot{\mathbf{D}}_k \succeq \mathbf{0}\}}{\text{minimize}} & \sum_{k=1}^{K} ||\dot{\mathbf{D}}_k - \tilde{\mathbf{D}}_k||_F^2 \\ \text{subject to} & \sum_{k=1}^{K} \text{tr}(\dot{\mathbf{D}}_k) = P. \end{array} \quad (6)$$

It is easy to see that $\dot{\mathbf{D}}_k$ must be diagonal to minimize the objective of (6). Next let $\bar{\mathbf{d}}_k = \mathrm{diag}(\bar{\mathbf{D}}_k)$, $\tilde{\mathbf{d}}_k = \mathrm{diag}(\tilde{\mathbf{D}}_k)$, $\bar{\mathbf{d}} = [\bar{\mathbf{d}}_1^T, \bar{\mathbf{d}}_2^T, \ldots, \bar{\mathbf{d}}_K^T]^T$, and $\tilde{\mathbf{d}} = [\tilde{\mathbf{d}}_1^T, \tilde{\mathbf{d}}_2^T, \ldots, \tilde{\mathbf{d}}_K^T]^T$. Then (6) can be reduced to

$$\begin{array}{ll} \underset{\bar{\mathbf{d}} \geq 0}{\text{minimize}} & \frac{1}{2}\|\bar{\mathbf{d}} - \tilde{\mathbf{d}}\|_2^2 \\ \text{subject to} & \mathbf{1}_{\tilde{M}} \bar{\mathbf{d}} = P \end{array} \quad (7)$$

where $\tilde{M} = \sum_1^K M_k$. It is now clear that (7) is the projection onto a canonical simplex and efficient algorithms (similar to water-filling algorithms) can be found in [17]. The complete description of the proposed CGP method for solving (4) is provided in Algorithm 1. We note that similar approaches were also presented in [10], [11].

---

**Algorithm 1:** The proposed CGP algorithm for solving (4).

**Input:** $P$, $\epsilon > 0$
1. Initialization: $\tau = 1 + \epsilon$, $m = 0$, $\{\bar{\mathbf{S}}_k^0\} \in \mathcal{S}$.
2. **while** $(\tau > \epsilon)$ **do**
3.     Calculate the conjugate gradient $\tilde{\mathbf{G}}_k^m$.
4.     Choose an appropriate positive scalar $s^m$ and create $\tilde{\mathbf{S}}^m = \bar{\mathbf{S}}_k^m + s^m \tilde{\mathbf{G}}_k^m$.
5.     Project $\tilde{\mathbf{S}}_k^m$ onto $\mathcal{S}$ to obtain $\dot{\mathbf{S}}_k^m$.
6.     Choose appropriate step size $\alpha^m$ and set $\bar{\mathbf{S}}_k^{m+1} = \bar{\mathbf{S}}_k^m + \alpha^m(\dot{\mathbf{S}}_k^m - \bar{\mathbf{S}}_k^m)$.
7.     $\tau = |\mathrm{tr}(\nabla f(\bar{\mathbf{S}}_k^m)^\dagger (\bar{\mathbf{S}}_k^{m+1} - \bar{\mathbf{S}}_k^m))|$.
8.     $m := m + 1$.
9. **end**

**Output:** $\bar{\mathbf{S}}_k$ as the optimal solution to (4).

---

Another main step of a CGP method is the computation of the conjugated gradient of the objective, as required in line 3 of Algorithm 1. The conjugate gradient, denoted as $\tilde{\mathbf{G}}_k^m$, can be calculated as follows. First, we compute the gradient of the objective in (4) as

$$\nabla f(\bar{\mathbf{S}}_k) = -\mathbf{H}_k \sum_{j=1}^k \Delta_j \left( \mathbf{Q} + \sum_{i=j}^K \mathbf{H}_i^\dagger \bar{\mathbf{S}}_i \mathbf{H}_i \right)^{-1} \mathbf{H}_k^\dagger. \quad (8)$$

Then the conjugate gradient direction is given by

$$\tilde{\mathbf{G}}_k^m = -\nabla f(\bar{\mathbf{S}}_k^m) + \beta^m \tilde{\mathbf{G}}_k^{m-1} \quad (9)$$

where the parameter $\beta^m$ is the Fletcher choice of deflection [18]

$$\beta^m = \begin{cases} 0 & m = 0 \\ \frac{-\|\nabla f(\bar{\mathbf{S}}_k^m)\|^2}{\mathrm{tr}((\tilde{\mathbf{G}}_k^{m-1})^\dagger \nabla f(\bar{\mathbf{S}}_k^{m-1}))} & m \geq 1 \end{cases} \quad (10)$$

For the step size in line 6 of Algorithm 1, we perform an Armijo line search [19] to determine appropriate value.

For the importance case of the sum capacity of the MIMO BC, we note that more efficient approaches to solving (4) do exist. In fact, in this case (4) becomes

$$\begin{array}{ll} \underset{\{\bar{\mathbf{S}}_k \succeq 0\}}{\text{maximize}} & \log|\mathbf{Q}^n + \sum_{i=1}^K \mathbf{H}_i^\dagger \bar{\mathbf{S}}_i \mathbf{H}_i| \\ \text{subject to} & \sum_{k=1}^K \mathrm{tr}(\bar{\mathbf{S}}_k) = P. \end{array} \quad (11)$$

We remark that (11) is equivalent to the problem of finding the sum-capacity of a MAC with a SPC for which the sum power iterative water-filling proposed in [20] or the dual decomposition method in [9] have been shown to be particularly computationally efficient.

We now turn our attention to the problem of finding $\mathbf{Q}^{n+1}$. If a pure AO method is followed, we arrive at the optimization problem below:

$$\begin{array}{ll} \underset{\mathbf{Q} \succeq 0}{\text{minimize}} & \sum_{k=1}^K \Delta_k \log|\mathbf{Q} + \sum_{i=k}^K \mathbf{H}_i^\dagger \bar{\mathbf{S}}_i \mathbf{H}_i| \\ & -w_K \log|\mathbf{Q}| \\ \text{subject to} & \mathrm{tr}(\mathbf{Q}\mathbf{P}) = P \end{array} \quad (12)$$

However, as mentioned in [14] and also observed in [16], the convergence of such a naive AO method is not guaranteed. The novelty of our proposed AO algorithm is that, instead of optimizing the original objective in (12) which can lead to fluctuations, we opt to minimize an upper bound of the objective in (12). This is in light of the SCA principle, and will lead to a monotonic convergence as shown in the Appendix. To this end, by invoking the concavity of $\log\det$ function, we have the following inequality

$$\log|\mathbf{Q} + \sum_{i=k}^K \mathbf{H}_i^\dagger \bar{\mathbf{S}}_i^n \mathbf{H}_i| \leq \log|\mathbf{\Phi}_k^n| + \mathrm{tr}\big(\mathbf{\Phi}_k^{-n}(\mathbf{Q} - \mathbf{Q}^n)\big) \quad (13)$$

where $\mathbf{\Phi}_k^n = \mathbf{Q}^n + \sum_{i=k}^K \mathbf{H}_i^\dagger \bar{\mathbf{S}}_i^n \mathbf{H}_i$, $\mathbf{\Phi}_k^{-n} \triangleq (\mathbf{\Phi}_k^n)^{-1}$. Thus, using the above upper bound, $\mathbf{Q}^{n+1}$ is found to be the optimal solution to the following problem

$$\begin{array}{ll} \underset{\mathbf{Q} \succeq 0}{\text{minimize}} & \sum_{k=1}^K \frac{\Delta_k}{w_K} \mathrm{tr}\big(\mathbf{\Phi}_k^{-n}\mathbf{Q}\big) - \log|\mathbf{Q}| \\ \text{subject to} & \mathrm{tr}(\mathbf{Q}\mathbf{P}) = P. \end{array} \quad (14)$$

Since $\mathbf{Q}$ in (14) is diagonal, i.e., $\mathbf{Q} = \mathrm{diag}(\mathbf{q})$, we can rewrite (14) as

$$\begin{array}{ll} \underset{\mathbf{q} > 0}{\text{minimize}} & \sum_{i=1}^N \sum_{k=1}^K \phi_{ki}^n q_i - \log q_i \\ \text{subject to} & \sum_{i=1}^N P_i q_i = P, \end{array} \quad (15)$$

where $\phi_{ki}^n = \left[\frac{\Delta_k}{w_K}\mathbf{\Phi}_k^{-n}\right]_{i,i}$, i.e., $\phi_{ki}^n$ is the $i$th diagonal element of $\frac{\Delta_k}{w_K}\mathbf{\Phi}_k^{-n}$. By setting the derivative of Lagrangian function of (15) to zero, we obtain

$$q_i = \frac{1}{\sum_{k=1}^K \phi_{ki}^n + \gamma P_i} \quad (16)$$

where $\gamma \geq 0$ is the solution of the equation

$$\sum_{i=1}^N \frac{P_i}{\sum_{k=1}^K \phi_{ki}^n + \gamma P_i} = P. \quad (17)$$

Denote $g(\gamma) = \sum_{i=1}^N \frac{P_i}{\sum_{k=1}^K \phi_{ki}^n + \gamma P_i} - P$. It's easy to see that $g(\gamma)$ is decreasing with $\gamma$. When $\gamma = 0$, since $\sum_{k=1}^K \phi_{ki}^n \leq q_i^{-1}$ we have

$$\sum_{i=1}^N \frac{P_i}{\sum_{k=1}^K \phi_{ki}^n} \geq \sum_{i=1}^N P_i q_i = P. \quad (18)$$

Therefore, (17) always has a unique solution, which can found efficiently, e.g., by the bisection or Newton method. The proposed algorithm based on AO is summarized in Algorithm 2.

The main point of Algorithm 2 is to eliminate the possible ping-pong effect of the obtained objective by the use of the inequality in (13). The convergence proof of Algorithm 2 is provided in the Appendix. Note that the error tolerance $\tau$ of Algorithm 2 is only computed for $n \geq 1$.

---

**Algorithm 2:** Proposed algorithm for the computation of the capacity region of a MIMO BC based on AO.

**Input:** $\mathbf{Q} := \mathbf{Q}^0$ diagonal matrix of positive elements, $\epsilon > 0$

1. Initialization: Set $n := 0$ and $\tau = 1 + \epsilon$.
2. **while** ($\tau > \epsilon$) **do**
3.     Solve (4) and denote the optimal solution by $\{\bar{\mathbf{S}}_k^n\}$
4.     For $n \geq 1$ compute
$\tau = |f^{\text{DPC}}(\mathbf{Q}^n, \{\bar{\mathbf{S}}_k^n\}) - f^{\text{DPC}}(\mathbf{Q}^{n-1}, \{\bar{\mathbf{S}}_k^{n-1}\})|$,
where $f^{\text{DPC}}(\cdot)$ denotes the objective in (3).
5.     For each $k$, compute
$\mathbf{\Phi}_k^{-n} = (\mathbf{Q}^n + \sum_{i=k}^K \mathbf{H}_i^\dagger \bar{\mathbf{S}}_i^n \mathbf{H}_i)^{-1}$.
6.     Solve (14) to find $\mathbf{Q}^{n+1}$.
7.     $n := n + 1$.
8. **end**

**Output:** Use the obtained $\{\bar{\mathbf{S}}_k^n\}_{k=1}^K$ and the BC-MAC transformation in [5] to find the optimal solution to (2).

---

## IV. COMPLEXITY ANALYSIS

In this section, we analyze the complexity estimation of the proposed algorithm in terms of the number of flops. The results of flop counting for typical matrix operations such as EVD, matrix inversion are taken from [21] and [22]. Moreover, we treat every complex operation as 6 real flops as considered in [22], [23]. In the complexity analysis presented below, we only consider the main operations of high complexity in the overall complexity.

### A. Complexity of Algorithm 2

Algorithm 2 has two key procedures, one for finding $\{\bar{\mathbf{S}}_k\}$ and one for finding $\mathbf{Q}$. Of all steps in Algorithm 1, the computation of the conjugate gradient direction given in (8) has the largest computation cost. Similarly, the complexity for finding $\mathbf{Q}$ is mostly due to the computation of $\mathbf{\Phi}_k^{-n}$ defined in (13). Therefore, the per-iteration complexity of Algorithm 2 is $\mathcal{O}(KN^3)$. That is, the complexity of the proposed algorithm increases linearly with the number of users.

### B. Complexity of the Barrier Interior-Point Method in [14]

The method presented in [14] is based on solving the KKT condition for (3). To find the gradient w.r.t each input covariance matrix for the logdet function, the complexity is $6N^3$. Then, to find a Newton direction, this algorithm still needs to solve a linear system of $M(M+1)/2$ unknowns, which requires $6(\frac{KM(M+1)}{2} + 3N)^3$ flops.[1] In total, the per-iteration complexity is $\mathcal{O}(K^2N^3 + K^3M^6)$.

[1] Here we assume $M_k = M$ for all $k$ to simplify the notation.

TABLE I
PER-ITERATION COMPLEXITY COMPARISON

| Algorithms | Per-iteration complexity |
|---|---|
| Algorithm 2 | $\mathcal{O}(KN^3)$ |
| Barrier interior-point method [14] | $\mathcal{O}(K^2N^3 + K^3M^6)$ |

The complexity comparison is summarized in Table I. We remark that the complexity of Algorithm 2 increases linearly with the number of users $K$, which was also achieved in [20] for the sum-capacity with SPC. Thus, Algorithm 2 scales much better than the barrier interior-point method in [14] with the number of users.

## V. NUMERICAL RESULTS

This section provides numerical results to verify the proposed algorithm. If not otherwise mentioned, the number of transmit antennas at the BS is set to $N = 5$ and the number of receive antennas at each user is set to $M_k = 2$ for all $k$.. The total power $P$ is varied from a low value to a high one to investigate the convergence rate of the proposed algorithm. The power constraint for each transmit antenna is set equally. The initial value $\mathbf{Q}^0$ in the proposed algorithm is set to the identity matrix. An error tolerance of $\epsilon = 10^{-6}$ is selected as the stopping criterion for the proposed algorithm.

In the first simulation, we plot the convergence of the proposed algorithm for different value of the total transmit power for a set of randomly generated channel realizations. As can be seen, Fig. 1 shows that the convergence of the proposed algorithm is strictly monotonic as proved in the Appendix. An interesting observation is that the number of iterations decreases when the total power increases.

As mentioned in the complexity analysis, the proposed algorithm has a desirable property, i.e., the per-iteration complexity increases linearly with the number of users $K$. This is an attractive property in the system of a large number of users which is usually the case for a massive MIMO setup. However, it is difficult, if not impossible, to analyze the convergence rate of the proposed algorithm with $K$ by analytical expressions. Instead, we study the convergence property of the proposed algorithm with $K$ by numerical experiments. For the purpose, we plot in Fig. 2 the average number of iterations required by the proposed algorithm to converge. As can be seen, the number of iterations for the proposed algorithm to converge is relatively insensitive to $K$. It is worth noting that for similar setups, the interior-point method in [14] requires at least 60 iterations to converge, while the proposed algorithm only takes 4 iterations even in the case of 50 users. This promising characteristic of the proposed algorithm makes it suitable for studying the capacity region of massive MIMO systems where the number of transmit antennas and/or the number of users can be very large. This point will be further elaborated in the next numerical experiment.

Taking the advantage that the proposed algorithm has low complexity, in the last numerical experiment we characterize the capacity region of a massive MIMO system with PAPC. In particular, we also consider achievable rate region of the well-known ZF scheme [24], [25]. The purpose is to understand the performance of ZF (which is thought to be sub-optimal) in comparison with the capacity achieving coding scheme under some realistic channel models. To this end we consider a simple urban scenario using WINNER II B1 channel model [26], where a base station, equipped with $N = 128$ antennas, is located at the center of the cell and single-antenna receivers are distributed randomly. The total power at the BS is $P = 46$ dBm and each antenna is subject to equal power constraint, i.e., $P_i = P/N$ for $i = 1, 2, \ldots, 128$. As can be seen clearly in Fig. 3, there is a remarkable gap between the achievable rate region of ZF and the capacity region, especially when the number of users increases. This basically implies that ZF is still far from optimal for a practical number of transmit antennas. Our observation opens research opportunities in the future to strike the balance between optimal performance by DPC and low-complexity by ZF.

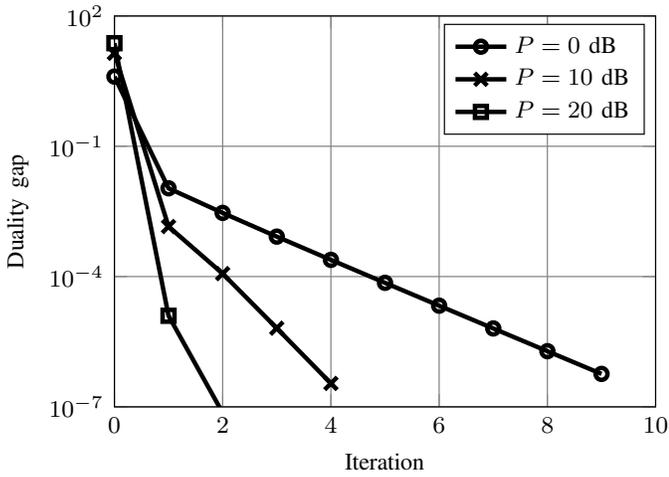

Fig. 1. Convergence behavior of the proposed algorithm with $K = 2$.

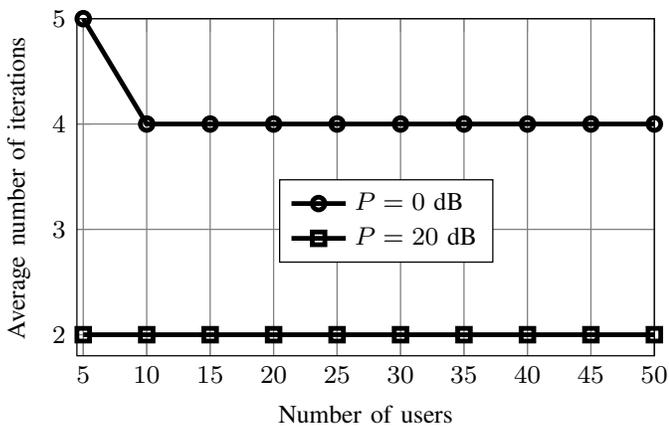

Fig. 2. Average number of iterations versus number of users.

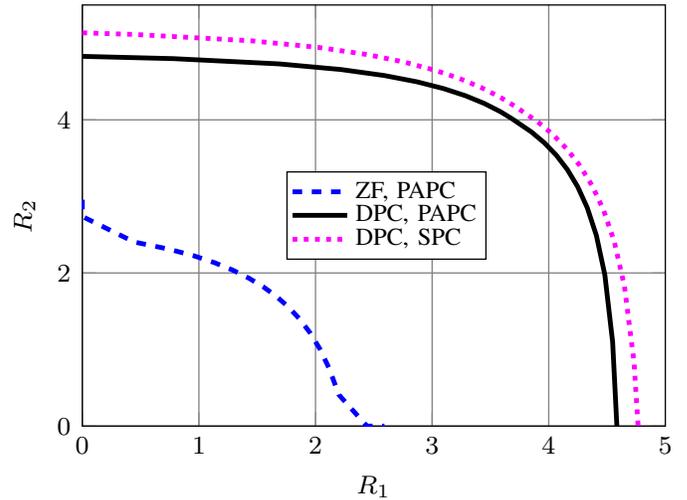

(a) Number of users $K = 2$

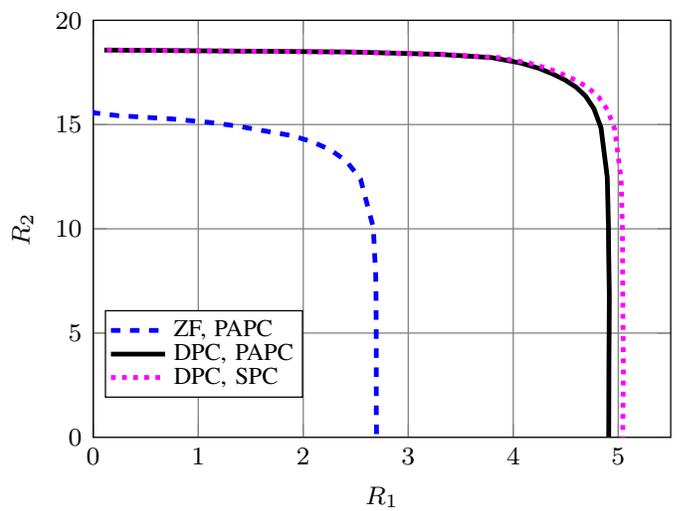

(b) Number of users $K = 8$

Fig. 3. Comparison of capacity region of a massive MIMO setup with $N = 128$ and $M_k = 1$ for all $k$. For the case $K = 8$ users, the capacity region is projected on the first two users.

## VI. CONCLUSIONS

In this paper, we have considered the problem of computing the capacity region of Gaussian MIMO BCs subject to PAPC. Towards this end, the problem of WSRMax with PAPC has been solved by a low-complexity algorithm. We have first converted the noncovex problem of MIMO BC into an equivalent minimax problem in the corresponding dual MAC. Then a novel AO algorithm has been proposed to solve the resulting mimimax program in combination with the successive convex optimization principle. In particular, all the computation in the proposed algorithm is based on closed-form expressions. In addition, the simulation results have demonstrated a fast and stable convergence of the proposed algorithm, even for large-scale settings.

## ACKNOWLEDGEMENTS

This work was supported by a research grant from Science Foundation Ireland and is co-funded by the European Regional Development Fund under Grant 13/RC/2077.

## APPENDIX

For notational convenience, we denote $\mathcal{Q} \triangleq \{\mathbf{Q}|\mathbf{Q} : \text{diagonal}, \mathbf{Q} \succeq \mathbf{0}, \text{tr}(\mathbf{QP}) = P\}, \hat{\Delta}_k = \frac{\Delta_k}{w_K}$. Since the function $\log|\mathbf{Q} + \sum_{i=k}^K \mathbf{H}_i^\dagger \bar{\mathbf{S}}_i \mathbf{H}_i|$ is *jointly concave* with $\mathbf{Q}$ and $\{\bar{\mathbf{S}}_k\}$, the following inequality holds

$$\sum_{k=1}^K \hat{\Delta}_k \log|\mathbf{Q} + \sum_{i=k}^K \mathbf{H}_i^\dagger \bar{\mathbf{S}}_i \mathbf{H}_i|$$
$$\leq \sum_{k=1}^K \hat{\Delta}_k \log|\underbrace{\mathbf{Q}^n + \sum_{i=k}^K \mathbf{H}_i^\dagger \bar{\mathbf{S}}_i^n \mathbf{H}_i}_{\boldsymbol{\Phi}_k^n}|$$
$$+ \sum_{k=1}^K \hat{\Delta}_k \text{tr}\big(\boldsymbol{\Phi}_k^{-n}(\mathbf{Q} - \mathbf{Q}^n)\big)$$
$$+ \sum_{k=1}^K \hat{\Delta}_k \sum_{i=k}^K \text{tr}(\mathbf{H}\boldsymbol{\Phi}_k^{-n}\mathbf{H}^\dagger(\bar{\mathbf{S}}_i - \bar{\mathbf{S}}_i^n)) \quad (19)$$

for all $\mathbf{Q} \in \mathcal{Q}$ and $\{\bar{\mathbf{S}}_k\} \in \mathcal{S}$. The above inequality comes from the first order approximation of $\log|\mathbf{Q} + \sum_{i=k}^K \mathbf{H}_i^\dagger \bar{\mathbf{S}}_i \mathbf{H}_i|$ around the point $(\mathbf{Q}^n, \{\bar{\mathbf{S}}_k^n\})$. Substitute $\mathbf{Q} := \mathbf{Q}^{n+1}$ and $\bar{\mathbf{S}}_k := \bar{\mathbf{S}}_k^{n+1}$ into the above equality, we have

$$\sum_{k=1}^K \hat{\Delta}_k \log|\mathbf{Q}^{n+1} + \sum_{i=k}^K \mathbf{H}_i^\dagger \bar{\mathbf{S}}_i^{n+1} \mathbf{H}_i|$$
$$\leq \sum_{k=1}^K \hat{\Delta}_k \Big(\log|\boldsymbol{\Phi}_k^n| + \text{tr}(\boldsymbol{\Phi}_k^{-n}(\mathbf{Q}^{n+1} - \mathbf{Q}^n))$$
$$+ \sum_{i=k}^K \text{tr}(\mathbf{H}\boldsymbol{\Phi}_k^{-n}\mathbf{H}^\dagger(\bar{\mathbf{S}}_i^{n+1} - \bar{\mathbf{S}}_i^n))\Big). \quad (20)$$

Since $\{\bar{\mathbf{S}}_k^n\} = \arg\max_{\{\bar{\mathbf{S}}_k\}\in\mathcal{S}} \sum_{k=1}^K \Delta_k \log|\mathbf{Q}^n + \sum_{i=k}^K \mathbf{H}_i^\dagger \bar{\mathbf{S}}_i \mathbf{H}_i|$, the optimality condition results in

$$\sum_{k=1}^K \hat{\Delta}_k \sum_{i=k}^K \text{tr}(\mathbf{H}\boldsymbol{\Phi}_k^{-n}\mathbf{H}^\dagger(\bar{\mathbf{S}}_i - \bar{\mathbf{S}}_i^n)) \leq 0 \quad (21)$$

for all $\{\bar{\mathbf{S}}_k\} \in \mathcal{S}$. For $\{\bar{\mathbf{S}}_k\} = \{\bar{\mathbf{S}}_k^{n+1}\}$ the above inequality is equal to

$$\sum_{k=1}^K \hat{\Delta}_k \sum_{i=k}^K \text{tr}(\mathbf{H}\boldsymbol{\Phi}_k^{-n}\mathbf{H}^\dagger(\bar{\mathbf{S}}_i^{n+1} - \bar{\mathbf{S}}_i^n)) \leq 0 \quad (22)$$

which leads to

$$\sum_{k=1}^K \hat{\Delta}_k \log|\mathbf{Q}^{n+1} + \sum_{i=k}^K \mathbf{H}_i^\dagger \bar{\mathbf{S}}_i^{n+1} \mathbf{H}_i|$$
$$\leq \sum_{k=1}^K \hat{\Delta}_k \Big(\log|\boldsymbol{\Phi}_k^n| + \text{tr}\big(\boldsymbol{\Phi}_k^{-n}(\mathbf{Q}^{n+1} - \mathbf{Q}^n)\big)\Big). \quad (23)$$

Subtract both sides of the above inequality by $\log|\mathbf{Q}^{n+1}|$ results in

$$\underbrace{\sum_{k=1}^K \hat{\Delta}_k \log|\mathbf{Q}^{n+1} + \sum_{i=k}^K \mathbf{H}_i^\dagger \bar{\mathbf{S}}_i^{n+1} \mathbf{H}_i| - \log|\mathbf{Q}^{n+1}|}_{f^{\text{DPC}}(\mathbf{Q}^{n+1}, \{\bar{\mathbf{S}}_k^{n+1}\})}$$
$$\leq \sum_{k=1}^K \hat{\Delta}_k \Big(\log|\boldsymbol{\Phi}_k^n| + \text{tr}\big(\boldsymbol{\Phi}_k^{-n}(\mathbf{Q}^{n+1} - \mathbf{Q}^n)\big)\Big) - \log|\mathbf{Q}^{n+1}|.$$
(24)

Since $\mathbf{Q}^{n+1}$ solves (14) it holds that

$$\sum_{k=1}^K \hat{\Delta}_k \Big(\log|\boldsymbol{\Phi}_k^n| + \text{tr}\big(\boldsymbol{\Phi}_k^{-n}(\mathbf{Q}^{n+1} - \mathbf{Q}^n)\big)\Big) - \log|\mathbf{Q}^{n+1}|$$
$$\leq \sum_{k=1}^K \hat{\Delta}_k \Big(\log|\boldsymbol{\Phi}_k^n| + \text{tr}\big(\boldsymbol{\Phi}_k^{-n}(\mathbf{Q} - \mathbf{Q}^n)\big)\Big) - \log|\mathbf{Q}| \quad (25)$$

for all $\mathbf{Q} \in \mathcal{Q}$. For the special case $\mathbf{Q} := \mathbf{Q}^n$, the inequality above is reduced to

$$\sum_{k=1}^K \hat{\Delta}_k \Big(\log|\boldsymbol{\Phi}_k^n| + \text{tr}\big(\boldsymbol{\Phi}_k^{-n}(\mathbf{Q}^{n+1} - \mathbf{Q}^n)\big)\Big) - \log|\mathbf{Q}^{n+1}|$$
$$\leq \underbrace{\sum_{k=1}^K \hat{\Delta}_k \log|\boldsymbol{\Phi}_k^n| - \log|\mathbf{Q}^n|}_{f^{\text{DPC}}(\mathbf{Q}^n, \{\bar{\mathbf{S}}_k^n\})}. \quad (26)$$

Combining (24) and (26) results in $f^{\text{DPC}}(\mathbf{Q}^n, \{\bar{\mathbf{S}}_k^n\}) \geq f^{\text{DPC}}(\mathbf{Q}^{n+1}, \{\bar{\mathbf{S}}_k^{n+1}\})$.

It is easy to see that $\{f^{\text{DPC}}(\mathbf{Q}^n, \{\bar{\mathbf{S}}_k^n\})\}$ is bounded above, and thus $\{f^{\text{DPC}}(\mathbf{Q}^n, \{\bar{\mathbf{S}}_k^n\})\}$ is convergent. We also note that (13) is strict if $\mathbf{Q} \neq \mathbf{Q}^n$. Consequently, the sequence $\{f^{\text{DPC}}(\mathbf{Q}^n, \{\bar{\mathbf{S}}_k^n\})\}$ is *strictly decreasing* unless it is convergent. Therefore, the continuity of $f^{\text{DPC}}(\cdot)$ and the compactness of $\mathcal{S}$ and $\mathcal{Q}$ imply $\lim_{n\to\infty} f^{\text{DPC}}(\mathbf{Q}^n, \{\bar{\mathbf{S}}_k^n\}) = f^{\text{DPC}}(\mathbf{Q}^*, \{\bar{\mathbf{S}}_k^*\})$.